\documentclass[%
apstemplate,
amsmath,amssymb,
reprint,
]{revtex4-2}

\usepackage{graphicx}
\usepackage{dcolumn}
\usepackage{bm}
\usepackage[utf8]{inputenc}
\usepackage[T1]{fontenc}
\usepackage{mathptmx}

\usepackage{amsmath}
\usepackage{hyperref}
\usepackage{tikz}

\usepackage{tikz}
\usepackage{pgfplots}
\usetikzlibrary{patterns}
\usetikzlibrary{plotmarks}
\usetikzlibrary{positioning}
\usetikzlibrary{external}
\usetikzlibrary{patterns}
\tikzset{external/system call={pdflatex \tikzexternalcheckshellescape -halt-on-error 
-interaction=batchmode -jobname "\image" "\texsource" && 
pdftops -eps "\image".pdf}}
\tikzsetexternalprefix{external_figs/}

\newcommand{\GaO}{$\beta-$Ga$_2$O$_3$ }

\begin{document}

\title[Enhanced phonon-drag by nanoscale design of homoepitaxial \hbox{$\beta$-Ga$_2$O$_3$}]{Enhanced phonon-drag by nanoscale design of homoepitaxial \hbox{$\beta$-Ga$_2$O$_3$}}

\author{Johannes Boy}
\affiliation{Novel Materials Group, Humboldt-Universität zu Berlin, Newtonstraße 15, 12489 Berlin, Germany}
\author{R\"udiger Mitdank}
\affiliation{Novel Materials Group, Humboldt-Universität zu Berlin, Newtonstraße 15, 12489 Berlin, Germany}
\author{Andreas Popp}
\affiliation{Leibniz-Institut f\"ur Kristallz\"uchtung, Max-Born-Strasse 2, 12489 Berlin, Germany}
\author{Zbigniew Galazka}
\affiliation{Leibniz-Institut f\"ur Kristallz\"uchtung, Max-Born-Strasse 2, 12489 Berlin, Germany}
\author{Saskia F. Fischer}
\email[]{contact author: sfischer@physik.hu-berlin.de}
\affiliation{Novel Materials Group, Humboldt-Universität zu Berlin, Newtonstraße 15, 12489 Berlin, Germany}
\affiliation{Center for the Sciences of Materials, Humboldt-Universität zu Berlin, Zum Großen Windkanal 2, 12489 Berlin, Germany}

\date{\today}

\begin{abstract}
Phonon drag may be harnessed for thermoelectric generators and devices.  Here, we demonstrate the geometric control of the phonon-drag contribution to the thermopower. In nanometer-thin electrically conducting $\beta$-Ga$_2$O$_3$ films homoepitaxially-grown on insulating substrates it is enhanced from -0,4 mV/K to up to -3 mV/K at 100 K by choice of the film thickness. Analysis of the temperature-dependent Seebeck coefficients reveal that a crossover from three-dimensional to quasi-two-dimensional electron-phonon interaction occurs for film thicknesses below 75~nm. The ratio of phonon-phonon to electron-phonon relaxation times in these confined structures is $10$ times larger than that of bulk. Generally the phonon drag can be tuned depending on the relations between the phonon-drag interaction length $\lambda_\text{PD}$, the phonon mean free path $\lambda$ and the film thickness $d$. Phonon drag can be enhanced for $\lambda_\text{PD}\gg\lambda>d$.  
\end{abstract}

\maketitle
\begin{figure}[hbtp]
\centering
\includegraphics[scale=0.6]{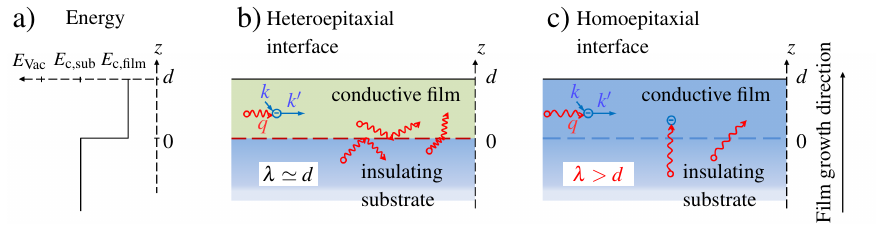}
\caption{\label{Struct} 
(a) Schematic drawing of the conduction band edge $E_\text{c}$ over film/substrate depth $z$. The electron mean free path is smaller than the film thickness $l_\text{e}<d$. (b) and (c) schematic drawing of the phonon-drag momentum transfer processes with phonons ($q$) and electrons ($k,k'$) in (b) heteroepitaxial films and (c) homoepitaxial films (this work). For any heterointerface the phonon mean free path $\lambda$ is reduced by  boundary scattering of the phonons at the heterointerface, whereas high-quality homoepitaxial films provide phonon-transparent interfaces.}
\end{figure}
\begin{figure}[hbtp]
\centering
\includegraphics[scale=0.8]{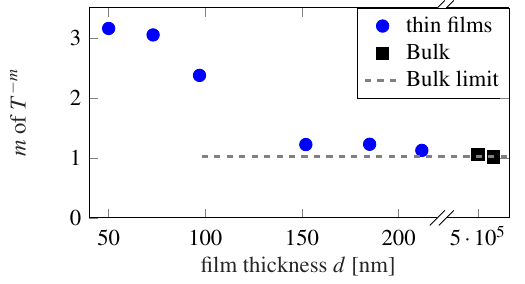}
\caption{\label{Parm} Crossover from three-dimensional to two-dimensional electron-phonon interaction occurs for thin films, resulting in giant phonon-drag. Exponent $m$ ($S_\text{PD}\propto T^{-m}$) as a function of film thickness $d$ for various $\beta$-Ga$_2$O$_3$ bulk crystals and thin films. A steady increase of $m$ with decreasing film thickness due to a change of phonon-transport in the thin film can be observed.}
\end{figure}

{\em Introduction--} Phonon drag - the process whereby phonons drag along charge carriers - leads to a contribution to the thermopower in use for energy harvesting, thermoelectric generators and low-noise applications. The phonon drag depends on the phonon-phonon and electron-phonon interactions~\cite{Herring1954}, and is influenced  by the material choice, microstructure and defects~\cite{TEH2005,Takahashi2016,Matsuura2019,Fauziah2020}. Further, it is influenced by the mean-free path of charge carriers $l_\text{e}$ and the dimensionality of the electrically conductive region~\cite{Yalamarthy2019,Pallecchi2016}. However, tailoring electron-phonon coupling by geometric control of homoepitaxial layers with {\em phonon-transparent} interfaces to the substrate~\cite{Ahrling2024} has not yet been elucidated.\\
Previous studies of phonon drag in confined structures have predominantly focused on heteroepitaxial structures. In a conducting film phonons drag charge carriers along by a transfer of momentum in-plane (Fig. \ref{Struct} (a) and (b)). The phonon drag is determined by the material choices of film and substrate, whereby both, phonons and charges, are scattered at the interface~\cite{Wang2013,Yalamarthy2019}. In particular, low-frequency acoustic phonons are effective contributors to the phonon drag because of their long lifetimes and, hence, long phonon mean free paths $\lambda$~\cite{TEH2005}. Phonon drag can be enhanced or decreased by confining the electrically conductive layer in polycrystalline~\cite{Kockert2019} and heteroepitaxial layers~\cite{Wang2013,Yalamarthy2019}. 
\\
Homoepitaxial structures, instead, as depicted schematically in Fig.~\ref{Struct} (c) may allow ballistic phonon transport within the confined electrically conductive layers and through the homo-epitaxial, i.e. phonon-transparent, interface into the single crystalline substrate. This allows to decouple the cross sections of electron-phonon and phonon-phonon interactions. Hence, by geometrical design it is expected that the phonon drag can be tuned in homoepitaxially grown layers.
\\
\\
In this Letter, we demonstrate phonon-drag enhancement by nanoscale design of homoepitaxial thin electrially conducting films $\beta$-Ga$_2$O$_3$ with phonon-transparent interfaces to the insulating substrate. The temperature-dependent thermoelectrical transport properties are investigated as a function of decreasing thickness, $d$, of the conducting film. The thermodiffusive~$S_\text{d}$ and phonon-drag~$S_\text{PD}$ contributions to the measured Seebeck coefficient~$S$ are analyzed with respect to a crossover from three-dimensional  to two-dimensional electron-phonon interaction. Electrically conducting thin homoepitaxial layers (by doping) are combined with a phononic reservoir of the insulating substrate (by compensation doping) barely limited by boundary scattering. The same order of doping density in the conducting layer and  compensation doping density insulating substrate ensure equivalent impurity scattering for phonons. 
\\ 
\begin{figure*}[hbtp]
\centering
\includegraphics[scale=1.0]{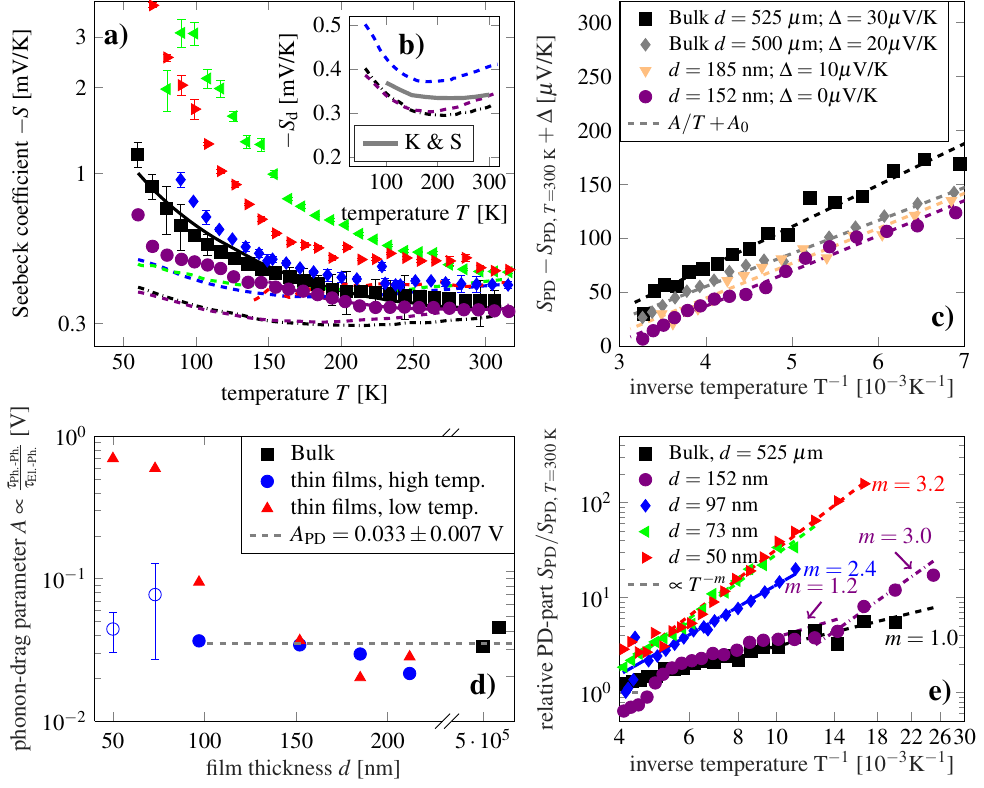}
\caption{\label{PhononDrag} 
(a) Experimental and theoretical Seebeck coefficient $S$ and (b) thermodiffusion part of the Seebeck coefficient including values from Kumar and Singisetti~\cite{Kumar2020} for $\beta$-Ga$_2$O$_3$ with an electron density of $n_\text{K\&S}=5.5\cdot10^{17}$ cm$^{-3}$ as a function of temperature $T$ for all \GaO bulk and thin films with film thickness $d$. In (a) the thermodiffusion part (dashed line) for all samples and thermodiffusion plus phonon-drag part (solid) for the bulk sample is additionally shown. The thermodiffusion part is discussed in the Supplementary Information. The same legend as in (e) applies. (c) Relative Phonon-drag part and fit $S_\text{PD}=A/T+A_0$ as a function of inverse temperature for various bulk crystals and thick ($d\geq150$ nm) films with offsets $\Delta$ for clarity. The $d=185$ nm sample has been reported before~\cite{Boy2019}. (d) Phonon-drag parameter $A$ as a function of film thickness $d$ for various $\beta$-Ga$_2$O$_3$ bulk crystals and thin films for low ($T\leq150$ K) and high ($T\geq150$ K) temperatures. The high temperature regime of the thinnest films is well described by $S_\text{d}$, hence there is a major computational uncertainty for the high temperature phonon-drag part. The uncertainty of the solid symbols is smaller than the symbol size. (e) Phonon-drag part normalized to the room temperature value as a function of inverse temperature for various bulk crystals and films. The lines represent fits with $S_\text{PD}\propto T^{-m}$.}
\end{figure*}
\\
{\em Material: $\beta$-Ga$_2$O$_3$-- } Of the wide bandgap ($E_\text{g}=4.85$ eV) semiconductor $\beta$-Ga$_2$O$_3$  homoepitaxial thin films and bulk single crystals of high structural quality are available~\cite{Galazka2016,Mazzolini2019,Cheng2018,Schewski2019,Anooz2020,Comstock2012,Galazka2021,Galazka2020}. Room temperature mobilities of $\mu>100$ cm$^2$/Vs and high power~\cite{Higashiwaki2018, Zhou2017, Yang2017}, optoelectronic~\cite{Galazka2018} and gas sensing~\cite{Manandhar2020, Afzal2019, Liu2008} device applications are reported. Previously, we determined the temperature-dependent electrical conductivity~\cite{Ahrling2019}, thermal conductivity ~\cite{Handwerg2015, Handwerg2016} and Seebeck coefficient ~\cite{Boy2019}.  Foundational for this study, we reported on ballistic phonon transport through effectively phonon-transparent interfaces in high-quality homo-epi film,s~\cite{Ahrling2024}, recently. In this work,  Si-doped semiconducting homo-epi films on Mg-compensation doped insulating substrates both with doping densities in the same order of magnitude of $10^{17}-10^{18}$~cm$^{-3}$ are used. Growth details, structural analysis and electrical characterization confirming the high quality of all samples are found in~\cite{SuppMat}.
\\
{\em Transport parameters--} The charge carrier density, mobility and Seebeck coeffient were determined for temperatures from 300 K to below 50 K \cite{SuppMat}. 
The Hall-charge carrier density shows comparable doping for all samples ($n_{T=300\text{ K}}=2.6\cdot10^{17}$ to $6.5\cdot10^{17}$ cm$^{-3}$) in magnitude and temperature dependence and reveals a donator level in the range of $E_{\text{D}_1}=18$ to $33$ meV.
The mobility is limited by polar optical phonons, neutral impurites and planar defects at high, intermediate and low temperatures, respectively. 
For very thin conducting epi-films $d\leq73$~nm on insulating substrates the scattering on planar defects increases and dominates the entire temperature regime.
The measured Seebeck coefficient $S$ as a function of temperature is depicted in Fig.~\ref{PhononDrag} (a) including the thermodiffusion part of the Seebeck coefficient determined from the reduced chemical potential and scattering parameter. The room temperature values for all samples are in the range $S_{T=300\text{ K}}=-330$ to $-460$ $\mu$V/K and increase with decreasing temperatures. The variety of the room temperature values for $d\geq73$~nm is due to different charge carrier concentrations. The higher Seebeck coefficient of the $d=50$~nm epi-film is additionally caused by the increase of the scattering factor, see~\cite{SuppMat}. 
\\
At cryogenic temperatures the Seebeck coefficient increases for all samples. However, for epi-films of lower thickness than 100 nm the Seebeck coefficient increases strongly with decreasing film thickness, see Fig.~\ref{PhononDrag} (a) . As will be detailed below, the phonon-drag is enhanced by the {\em geometric} confinement of charge carriers in the thin conducting epi-films, whereas phonons may travel quasi-ballistically without boundary scattering into the insulating substrate.
\\
{\em Thermodiffusion--} First, the thermodiffusive part of the Seebeck coefficient is discussed. It is calculated from the reduced chemical potential and scattering parameter for the samples with the highest and lowest charge carrier density (Fig. \ref{PhononDrag} (b)). For comparison, the first-principle calculations by Kumar and Singisetti~\cite{Kumar2020} calculated for $\beta$-Ga$_2$O$_3$ with an electron density of $n_\text{K\&S}=5.5\cdot10^{17}$ cm$^{-3}$ are shown. The experimentally determined and calculated values are in agreement. The slight offsets between the different samples are due to differences in doping, which affects the reduced chemical potential and the thermodiffusion part depends on the reduced chemical potential $\eta$ and the scattering factor $r$ as detailed in \cite{SuppMat}. However, the thermodiffusion part of the Seebeck coefficient only accounts for a small subset  ($-300$ to $-500$ $\mu$V/K) of  $S$ at lower temperatures (Fig. \ref{PhononDrag} (b)).
\\ 
{\em Phonon drag--} The increase in the Seebeck coeffient observed at low temperatures for all samples is attributed to the phonon-drag. A systematic dependence of the phonon drag~$S_\text{PD}$ as a function of thickness is found after subtraction of the thermodiffusion part~$S_\text{d}$ from the measured total Seebeck coefficient~$S$. For homo-epi films of a thickness larger than 100 nm the values for the phonon drag part $S_\text{PD}(T)$ are independent of the film thickness~$d$. For homo-epi films of thickness less than 100 nm the slope of the phonon drag part $S_\text{PD}(T)$ increases as described by Figs. \ref{PhononDrag} (e) and \ref{Parm}.
\\
The  observed phonon-drag increase is analysed in the following. Generally, the absolute value of the phonon drag contribution is limited by the phonon occupation at ultra-low temperatures and by phonon scattering at high temperatures. In between it exhibits a maximum value. In this study, the phonon drag is measured at temperatures above this maximum, where the phonon transport is mainly determined by Umklapp scattering and the phonon mean free path is given by $\lambda\propto T^{-1}$. Accordingly, in this temperature regime the phonon-drag contribution to the Seebeck coefficient is determined by $S_\text{PD}=A/T+A_0$ ~\cite{Herring1954}. For the conducting bulk single crystal a fit including the thermodiffusion and phonon-drag contributions to the Seebeck coefficient agrees well with the measured data (black solid line in Fig. \ref{PhononDrag}~(a)). Slight remaining discrepancies can be explained by computational uncertainties of reduced chemical potential $\eta$ and scattering parameter $r$, and as well deviations from $S_\text{PD}$ to the approximation given by $A/T+A_0$. 
\\
Fig. \ref{PhononDrag}~(c) confirms for single crystal bulk and thick epi-films ($d\geq150$ nm) that $S_\text{PD}$ is  proportional to the inverse temperature. Compared to the conducting bulk single crystal, the thick epi-films show a small decrease of the phonon-drag, which originates from the slight increase of phonon scattering in the epitaxial films, due to the silicon doping and lattice imperfections, which leads to a reduction of the phonon lifetime $\tau_\text{Ph.-Ph.}$. From the fit to the data with \hbox{$S_\text{PD}=A/T+A_0$} for temperatures $T\geq150$ K, a phonon-drag parameter \hbox{$A=\text{const.}=(0.033\pm0.007)$ V} for bulk and thin films is determined.\\
A similar analysis for the thin films ($d\leq100$ nm), however, separating the low ($T\leq150$ K) and high ($T\geq150$ K) temperature regimes, is depicted in Fig. \ref{PhononDrag}~(d) and (e), respectively. A strong increase in the phonon-drag parameter occurs for film thicknesses below 75~nm, for  $A=(0.69\pm0.02)$~V.
\\
To understand the origin of the observed increase of $S_\text{PD}$ for thin films, note that it relates to the phonon-phonon scattering time $\tau_\text{Ph.-Ph.}$ and the electron-phonon scattering time $\tau_\text{El.-Ph.}$ according to~\cite{Herring1954,Hutson1961,Smith1990}
\begin{equation}
S_\text{PD}=-\frac{\nu^2}{T}\frac{1}{\mu_\text{AP.}}\tau_\text{Ph.-Ph.}=\frac{m^*\nu^2}{eT}\cdot\frac{\tau_\text{Ph.-Ph.}}{\tau_\text{El.-Ph.}}
\end{equation}
with $\mu_\text{AP.}$ denoting the total electron mobility from scattering mechanisms transferring crystal momentum from acoustic phonons to electrons, $\nu$ the sound velocity and $m^*$ the effective mass. Here, low-frequency acoustic phonons are considered mainly responsible for the effective momentum transfer in the phonon-drag~\cite{Herring1954}. Thus, the phonon-drag parameter $A$ directly depends on the ratio of the phonon-phonon to electron-phonon scattering times  
\begin{equation}\label{eq:A}
a\cdot A=\frac{\tau_\text{Ph.-Ph.}}{\tau_\text{El.-Ph.}},
\end{equation}
with $a=e/\left(m^*v^2\right)$. This relates to the observation in Fig. \ref{PhononDrag} (d) that \hbox{$A=T\cdot S_\text{PD}\propto\frac{\tau_\text{Ph.-Ph.}}{\tau_\text{El.-Ph.}}$} increases with decreasing film thickness for low temperatures at which Umklapp scattering is reduced in the electrically conducting layer.\\
While both scattering times $\tau_\text{Ph.-Ph.}$ and $\tau_\text{El.-Ph.}$ increase with decreasing temperature due to a reduction of occupied phonon modes and the decrease of average phonon energy, however, the increase of $A$ with decreasing film thickness $d$ implies that the ratio of the scattering times is subject to the decreasing effective cross sectional area of the relevant momentum transfer for the phonon-drag.\\
The phonon-drag interaction length  $\lambda_\text{PD}=v\cdot\tau_\text{Ph.-Ph.}=v\cdot a\cdot A\cdot\tau_\text{El.-Ph.}$ is directly proportional to the phonon-drag parameter $A$ and the electron-phonon scattering time $\tau_\text{El.-Ph.}$. In a first approximation, the scattering times $\tau_\text{El.-Ph.}$ for 300 K and 100 K are about $9.4\cdot10^{-13}$ s and $4.8\cdot10^{-12}$ s, respectively. With $A=0.033$~V and $a=1.46\cdot10^{4}$~1/V (with $m^*=0.313\;m_\text{e}$ and $v=6200$~m/s) the value $a\cdot A\approx500$ for bulk and thick films at 300 K and 100 K.
However, for the thinnest films, the phonon-drag parameter strongly increases, leading to $a\cdot A\approx10200$ at 100 K. Therefore, the phonon-drag interaction length increases from  $\lambda_\text{PD}(100\text{ K})\approx14\;\mu$m for bulk and thick films to $\lambda_\text{PD, thin film}(100\text{ K})\approx300\;\mu$m for thin films. $\lambda_\text{PD}$ thereby exceeds the phonon mean free path $\lambda(300\text{ K})\approx3$ nm to $\lambda(100\text{ K})\approx65$ nm by orders of magnitude~\cite{Handwerg2015}. This is due to the fact that the phonon drag is maximized if normal phonon scattering is dominant and decreases if Umklapp scattering occurs.\\
Generally, the electron-phonon coupling responsible for the phonon-drag occurs in all three dimensions in the bulk. However, in high-quality homoepitaxial films the relation between the film thickness $d$ to the phonon mean free path (MFP) $\lambda$ and the phonon-drag interaction length $\lambda_\text{PD}$ may determine the magnitude of the phonon-drag.  If the phonon mean free path exceeds the film thickness $\lambda>d$, the phonon-phonon Umklapp scattering in the electronically active region is reduced. Therefore, the low frequency phonons interact with the carriers undisturbed and hence, the phonon-drag contribution to the thermopower increases (Fig. \ref{Struct} (c)). 
This in-plane momentum of the charge carriers is mainly dominant when the phonon mean free path $\lambda$ describing the heat transfer exceeds the film thickness $d$. If $\lambda_\text{PD}\gg d>\lambda$, there is diffusive heat transport and the carriers can be scattered out of the plane by interaction with phonons. In the case of $\lambda_\text{PD}\gg\lambda>d$, the heat transfer vertical to the plane of the conducting layer is \textit{quasi-ballistic}. \\
Phonon-phonon and electron-phonon interaction take place predominantly in-plane. Here, the condition $\lambda_\text{PD}\gg\lambda>d$ is fulfilled for $d < 100$ nm at 100 K, leading to an enhancement of the phonon-drag due to the crossover to quasi two-dimensional phonon transport. From Fauziah \textit{et al}.~\cite{Fauziah2020} it is evident that the long wavelength phonons are responsible for the in-plane phonon drag. These phonons preferably propagate specularly. Furthermore, short wavelength phonons that contribute to the thermal conductivity by Umklapp scattering do not contribute to the phonon-drag. This means that local peculiarities with a diameter of the order of the surface roughness, such as lattice defects or impurities, are unessential for the interaction of electrons with phonons for the drag effect. In this study, the mean free path $\lambda_\text{PD}$ exceeds by far the surface roughness and therefore the specularly reflected lattice waves determine the phonon-drag. Therefore,  $S_\text{PD}$ coincides for thin films. This is contrary to  $S_\text{d}$ because there the interaction with boundary defects cannot be neglected.\\
{\em Dimensonality of the relevant phonon transport--}
The observed dimensional crossover of the phonon drag with decreasing film thickness, is analysed  to account for the nature of the phonon-phonon interaction and occupied phonon modes using~\cite{Herring1958,Herring1958b}
\begin{equation}\label{eq:m}
S_\text{PD}\propto T^{-7/2}\cdot T^{s/2+\gamma}.
\end{equation}
For longitudinal acoustic phonons with the phonon wave-vector $q\rightarrow0$ the exponent $s\rightarrow0$ and $\gamma=0$ and $\gamma=2$ at low and high temperatures, respectively. For transversal acoustic phonons with $q\rightarrow0$ the exponent  $s\rightarrow-1$ and $\gamma=0$ and $\gamma=3$ at low and high temperatures, respectively. Using the experimental data, the resulting temperature dependence of the phonon-drag part normalized to the room temperature value is shown in Fig. \ref{PhononDrag} (e). 
For lower $T$ the exponent $m$ approximates to 3 (Fig. \ref{PhononDrag} (e)) if normal scattering dominates over Umklapp scattering.
The total exponent $m=7/2-s/2-\gamma$, as derived from eq. \eqref{eq:m}, is shown in Fig. \ref{Parm} as a function of film thickness.  A change of $\Delta m=2$ between low and high temperature for longitudinal phonons occurs.
From these results it follows that the bulk  exhibits the temperature dependence $S_\text{PD}\propto T^{-m}$ with $m=1$.
This is expected because for bulk, in our case the thickness $d=525$ $\mu$m$\gg\lambda(300\text{ K})\approx3$ nm~\cite{Handwerg2015,Handwerg2016} is much larger than the phonon mean free path. 
Thus, Umklapp processes are dominant~\cite{Handwerg2015,Handwerg2016}, which explain the observed $m=1$.
Lowering the temperature to $T=100$ K the phonon mean free path increases to $\lambda\approx65$ nm. Hence, only for the thickest films $d\geq 150$ nm will $m=1$ for $T>100$ K (Fig. \ref{Parm}).
\\
For the thin films with $d\leq100$ nm it is determined that $m\rightarrow3$ for all temperatures. Here, $\lambda$ is comparable to the thickness $d$ of the epitaxial layers. This demonstrates a change of $\Delta m=2$ induced by geometric confinement of the electrically conductive layer from $d>150$ nm ($d>\lambda$) to $d\leq100$ nm ($d<\lambda$).
\\
Previously, the temperature induced change of the exponent $\Delta m=2$ for bulk with $d\gg\lambda$ was discussed by Herring~\cite{Herring1958,Herring1958b} for germanium single crystals. In elastic anisotropic crystals the exponent $m$ changes by $\Delta m=2$ if the temperature is sufficiently low, because high-energy scattering processes like Umklapp scattering do not annihilate phonons. In this study, Fig. \ref{Parm}, $\Delta m=2$ is induced by geometric confinement, which causes a reduced Umklapp scattering in the electrically conductive layer at low temperatures and enhances the phonon-drag contribution.
This result of geometric confinement holds generally true if $\lambda>d$ and $\lambda_\text{PD}~\gg~d$ ~\cite{Kockert2019,Kockert2019a} then the phonon drag can be enhanced~\cite{Ramayya2012,Sadhu2015}. 
\\
As shown, the phonon drag can be strongly enhanced by geometric confinement of conductive layer thickness below the phonon mean free path in homoepitaxial systems. The scattering cross sections of electron-phonon and phonon-phonon interactions can by tuned on the basis of phonon-transparent interfaces. The enhancement of the phonon-drag 
scales with the ratio of the phonon mean free path $\lambda$ and the film thickness $d$. For $\lambda/d\geq1$ a quasi two-dimensional phonon transport takes place in-plane of the electrically conductive film. In the presented case the Seebeck coefficient at room temperature nearly doubles due to the reduced Umklapp scattering (see Fig. \ref{PhononDrag} (a)).
\\
\begin{figure}[h]
\centering
\includegraphics[scale=0.8]{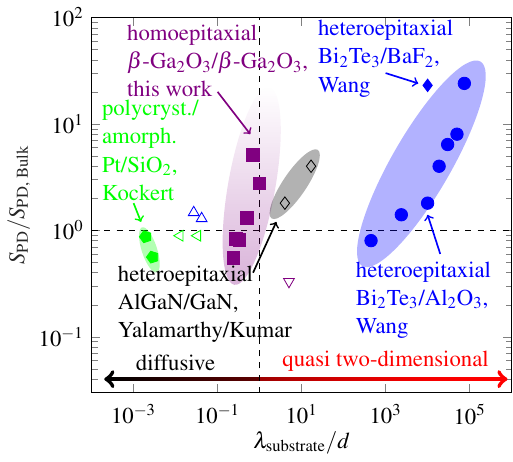}
\caption{\label{Concl}  Change of the phonon-drag part $S_\text{PD}/S_\text{PD, Bulk}$ over the ratio of the phonon mean free path of the substrate relative to the thickness of the electrically conductive material $\lambda_\text{substrate}/d$ for various quasi two-dimensional and quasi one-dimensional material systems. Shown is data for quasi two-dimensional systems of this work at $T=100$ K (violet filled squares), epitaxial Bi$_2$Te$_3$ on BaF$_2$ at $T=14$ K (blue filled diamond) and on Al$_2$O$_3$ at $T=30$ K (blue filled circle) by Wang~\cite{Wang2013}, sputtered and annealed Pt on SiO$_2$ at $T=70$ K (green filled pentagons) by Kockert~\cite{Kockert2019} and a two-dimensional electron gas in a AlGaN/GaN heterostructure by Yalamarthy~\cite{Yalamarthy2019} compared to data from Kumar~\cite{Kumar2021} at $T=100$ K (black unfilled diamonds). Data for quasi one-dimensional systems are shown for Si-nanowires at $T=100$ K (violet unfilled triangle pointing down) by Sadhu~\cite{Sadhu2015}, Ag- (blue unfilled triangles pointing up) and Bi- (green unfilled triangles pointing left) nanowires by Kockert~\cite{Kockert2019a,Kockert2021} at $T=75$ K and $T=130$ K, respectively.}
\end{figure}
\\
In Fig. \ref{Concl} a summary of the variations reached by tuning the phonon drag on behalf of geometrical confinement in various material systems including heteroepitaxial films (Bi$_2$Te$_3$/BaF$_2$~\cite{Wang2013}, Bi$_2$Te$_3$/Al$_2$O$_3$~\cite{Wang2013}, AlGaN/GaN~\cite{Yalamarthy2019}). 
Evidently, for systems with heteroepitaxial interfaces a higher $\lambda/d$ ratio is required to achieve the same increase of the phonon drag as for the homo-epi films from this study. In polycrystalline/amorphous metallic films of platinum on SiO$_2$~\cite{Kockert2019}  heat flow is dominated by electrons in the metal layer. $\lambda$ is very low in Pt and SiO$_2$ due to increased grain boundary scattering, lowering the phonon-drag. Finally, in nanowires ~\cite{Sadhu2015,Kockert2019a,Kockert2021}, a reduction of $\lambda$ due to boundary scattering on the surface of the nanowire was observed, which strongly reduces the phonon-drag effect.
\\
{\em Conclusion--} Phonon-drag engineering may be utilized for low-temperature thermo-electrical applications. Phonon drag can be either maximized by reducing phonon Umklapp scattering, lattice mismatch and boundary scattering or minimized vice versa. Nanoscale design of homoepitaxial films allow to select diffusive or quasi-ballistic heat transfer: confinement of charge carriers combined with phonon-transparent interfaces may pave the way to adjust cross-sections of electrons with phonons to obtain the desired performance and function.
\\
\\
{\em Data availability --} The data that support the findings of this study are available from the corresponding authors upon request.
\\
\\
{\em Acknowledgements --}
This work funded by the German Science Foundation (DFG FI932/10-2, DFG-FI932/10-1, DFG-FI932/11-1, WA 1453/3-1 and PO 2659/1-2) and was performed in the framework of GraFOx, a Leibniz-ScienceCampus partially funded by the Leibniz association. JB, RM and SFF are grateful for scientific and technical support by Dr. Olivio Chiatti.
\\
\\
{\em Competing interests --}
The authors declare no competing interests.\\

\nocite{*}
\bibliography{mybibfile_comb}

\onecolumngrid
\section*{Endmatter}
\twocolumngrid

{\em Single crystal growth--}
Bulk single crystals were grown by the Czochralski method ~\cite{Galazka2010,Galazka2014,Galazka2016,Galazka2021,Galazka2020}.
Substrates for homoepitaxy of 5x5 mm$^2$ were fabricated from the insulating Mg-compensation doped crystals.
\\
{\em Homoepitaxial film growth--}
Metal-organic vapour phase epitaxy was used for homo-epi film growth on (100) oriented substrates from single crystal growth. 
The substrates were off-oriented by 4$^\circ$ in [00-1] direction to avoid the formation of twin lamellae and to adjust the effective diffusion length to the terracce width with the aim to achieve step flow growth mode.
The thin films were Si-doped using Tetraethylorthosilicate to enhance n-type conductivity. 
More details  can be found elsewhere~\cite{Anooz2019,Anooz2020,Anooz2020a}.
Atomic force microscope (AFM) measurements were performed on all thin films after growth to confirm step-flow growth by the surface topography and low RMS (RMS$\leq1$~nm), see \cite{SuppMat}. 
\\
{\em Microlab for thermopower measurements--} 
The microlabs consisting of Ti/Au (7 nm/35 nm) metal lines on the surface of the films were manufactured by standard photolithography and magnetron sputtering. Deposited metal lines form a Schottky contact with the $\beta$-Ga$_2$O$_3$ and were contacted by attaching gold wire with indium at the contact-pads.
Ohmic contacts were manufactured by wedge bonding with an Al/Si-wire (99\%/1\%).
More details on the microlab and preparation can be found elsewhere~\cite{Ahrling2019,Boy2019} and details for this study in \cite{SuppMat}.
To ensure the highest quality of the microlab, quality control of the structure was checked by a Raith PIONEER Two scanning electron microscope system.
\\
{\em Measurement techniques--}
All measurements were performed using standard source-meter-units (SMU). 
The measurement of the current-voltage characteristics for the van-der-Pauw, Hall and Seebeck measurements were performed with a Keithley 2450 SMU.
The imprinting of the heating currents was done by a Keithley 2400 SMU.
All measurements were performed in a CryoVac Konti-IT flow cryostat.
The samples are in helium atmosphere at standard pressure. Van-der-Pauw- and Hall-measurements are detailed in \cite{SuppMat} 
\\
{\em Models for the electrical properties-- }
Application of fits of the charge neutrality equation and details of the scattering mechanisms that were used for determination of scattering rates are
detailed in \cite{SuppMat}.
\\
{\em Thermopower measurements--}
For the determination of the Seebeck-coefficients measurements are carried out using a microlab on the chip~\cite{Boy2019,Boy2020} consisting of metal line heaters, four-point metal line thermoemeters and ohmic contacts on these thermometers.
An electrical current is imprinted into one of the line heaters close to the edge of the sample, which creates Joule heating.
Thereon a temperature difference between the line heater and the edges of the sample is created which takes some time to become stable.
The emerging thermovoltage can be measured at the points of the thermometers where ohmic contacts were prepared.
At these points the temperature can be measured by the change of the resistance of the four-point resistance of the thermometer lines.
This procedure is done for several heating currents to calculate the Seebeck coefficient from the change of thermovoltage with temperature difference.
This method has been shown to be successful~~\cite{Boy2019,Boy2020} and carried out for various bulk and thin film samples at different temperatures in a flow cryostat between $T=50$ and $320$ K.
Below $T=50$ K the thermal conductivity of $\beta$-Ga$_2$O$_3$ limits the measurement of the thermovoltage in this setup.
\\
{\em Models for the thermoelectric properties--}
The reduced chemical potential $\eta=\left(E_\text{F}-E_\text{C}\right)/k_\text{B}T$ was calculated using the formula of Nilsson~\cite{Nilsson1973}, which is applicable to non-degenerate and degenerate semiconductors alike, and 
the calculation of the scattering parameter $r$ using the analytic expressions for the momentum relaxation time $\tau$, which were used to calculate the mobility, are detailed in \cite{SuppMat}.
\end{document}